\documentclass[doublecol]{epl2}

\title{Random Field XY Model in Three Dimensions: \\ The Role of Vortices}
\shorttitle{Random Field XY Model}

\author{D. A. Garanin \and E. M. Chudnovsky \and T. Proctor}
\shortauthor{Garanin \etal}

\institute{
  Physics Department, Lehman College, City University of New York, 250 Bedford Park Boulevard West, Bronx, New York 10468-1589, U.S.A.\\
 }

\abstract{
We study vortex states in a $3d$ random-field $xy$
model of up to one billion lattice
spins. Starting with random spin orientations, the sample freezes
into the vortex-glass state with a stretched-exponential decay of
spin correlations, having short correlation length and a low susceptibility, compared to vortex-free states.
In a field opposite to the initial magnetization, peculiar topological objects -- walls of spins still opposite
to the field -- emerge along the hysteresis curve. On increasing
the field strength, the walls develop cracks bounded by vortex loops. The
loops then grow in size and eat the walls away. Applications to magnets and superconductors are discussed.
}

\begin{document}

\maketitle

The $xy$ model with randomness is relevant to a number of physical
systems including magnets, superconductors, Josephson junction
arrays, spin- and charge-density waves. The question of interest
for superconductors is the distortion of the vortex lattice due to
the collective pinning of vortex lines by randomly distributed
point defects. In magnets it is a question of long-range behavior
of ferromagnetic correlations in the presence of torques applied
to individual spins by randomly distributed static local fields.
The order parameter in the flux lattice model is $\exp(iGu)$ with $u$
being the displacement and $G$ being a reciprocal lattice vector.
This makes  angle $\phi$ of the spin in the $xy$ model equivalent
to the displacement in the flux lattice.
The generic spin Hamiltonian is
\begin{equation}\label{Ham-descrete}
{\cal{H}} = -\frac{1}{2}\sum_{ij}J_{ij}{\bf s}_i \cdot {\bf s}_j
- \sum_i {\bf h}_i \cdot {\bf s}_i - {\bf H} \cdot \sum_i {\bf
s}_i,
\end{equation}
where 1/2 is compensating for the double counting of bonds in the exchange interaction between
nearest-neighbor ($J_{ij} \equiv J$) $xy$ spins $s$ in a cubic
lattice of spacing $a=1$. Other terms describe Zeeman
interaction of spins with the on-site random field ${\bf h}_i$
and the external field ${\bf H}$.
The continuous
counterpart of this model is
\begin{equation}\label{Ham-continuous}
{\cal{H}} = s \int d^3 r \left[\frac{Js}{2}({ \nabla}\phi)^2
- h \cos(\phi - \varphi) - H \cos\phi \right],
\end{equation}
where ${\bf r} = (x,y,z)$ is in units of $a$, $\phi({\bf r})$ is a
scalar field ($0 < \phi < 2\pi)$) that determines orientation of
the spin-field in the $xy$ plane, ${\bf s}({\bf r}) = s[\cos
\phi({\bf r}), \sin \phi({\bf r})]$, and $\varphi({\bf r})$
determines local orientation of the random field in the $xy$ plane:
${\bf h}({\bf r}) = h[\cos \varphi({\bf r}), \sin \varphi({\bf
r})]$, correlated as $\langle h_{\alpha}({\bf r}') h_{\beta}({\bf r}'')
\rangle = (h^2/2)\delta_{\alpha \beta}\delta({\bf r}'-{\bf
r}'')$.

In spite of the simplicity of the above model, the states
generated by it have not been well understood. The subject has a
long history. More than forty years ago Lar\-kin
\cite{Larkin-JETP1970} argued that a whatever weak random pinning
would destroy the long-range translational order in the
Abrikosov vortex lattice. Imry and Ma \cite{Imry-Ma-PRL1975}
generalized this statement for the case of any continuous-symmetry
magnetic order parameter in less than four dimensions. This implies that in
$3d$ the long-range ferromagnetic order would not
survive even a very weak static random field. The argument goes
like this. Smooth rotation of the magnetization on the scale $R$
costs exchange energy per spin of order $Js^2/R^2$. If ${\bf
h}({\bf r})$ was directed everywhere along the spin field, the
gain in the Zeeman energy per spin would be just $hs$. In reality,
however, the spin field can only follow average ${\bf h}({\bf r})$
so that the gain in the Zeeman energy per spin would scale as
$hs/R^{3/2}$ in $3d$. Minimization of the sum of the exchange and
Zeeman energies then gives the average size of Larkin-Imry-Ma
(LIM) domains, $R \sim (Js/h)^2$ at small $h$, implying zero total
magnetization in the absence of the external field.

More rigorously, at $H = 0$ the minimum of Eq. (\ref{Ham-continuous})
 satisfies
$\nabla^2 \phi = h_x\sin \phi - h_y \cos \phi$. Using the implicit
solution
\begin{equation}\label{angles}
\phi({\bf r}) = \frac{1}{Js}\int \frac{d^3r'}{4\pi |{\bf r} - {\bf
r}'|} [h_y({\bf r}')\cos \phi({\bf r}') - h_x({\bf r}') \sin \phi
({\bf r}')]
\end{equation}
 and averaging over the random field, one obtains
\begin{equation}\label{LIM-Rf}
\langle[\phi({\bf r}_1) - \phi({\bf r}_2)]^2\rangle = \frac{2|{\bf r}_1
- {\bf r}_2|}{R_f}, \quad R_f = 16\pi \left(\frac{Js}{h}\right)^2
\end{equation}
which, in the absence of vortices, is a rigorous result for $R \leq R_f$, as confirmed by our numerical studies. It was argued in
early works that the spin-spin
correlation function for any distance is given by
\begin{equation}\label{CF}
\langle{\bf
s}({\bf r}_1)\cdot{\bf s}({\bf r}_2)\rangle =
s^2 e^{-\langle[\phi({\bf r}_1) - \phi({\bf r}_2)]^2\rangle/2} = s^2 e^{-|{\bf r}_1 - {\bf r}_2|/R_f}.
\end{equation}
Effects of random magnetic anisotropy relevant to properties of
amorphous and sintered ferromagnets have been shown to resemble
those of random field \cite{RA}. Aizenman and Wehr
\cite{Aizenman-Wehr} provided a mathematical proof of the LIM
conjecture about the destruction of the long-range order by
quenched randomness. Early results on magnets and superconductors
have been summarized in Refs. \cite{Fisher-PRB1985} and
\cite{Blatter-RMP1994}.

\begin{figure}
\centering\includegraphics[width=8cm]{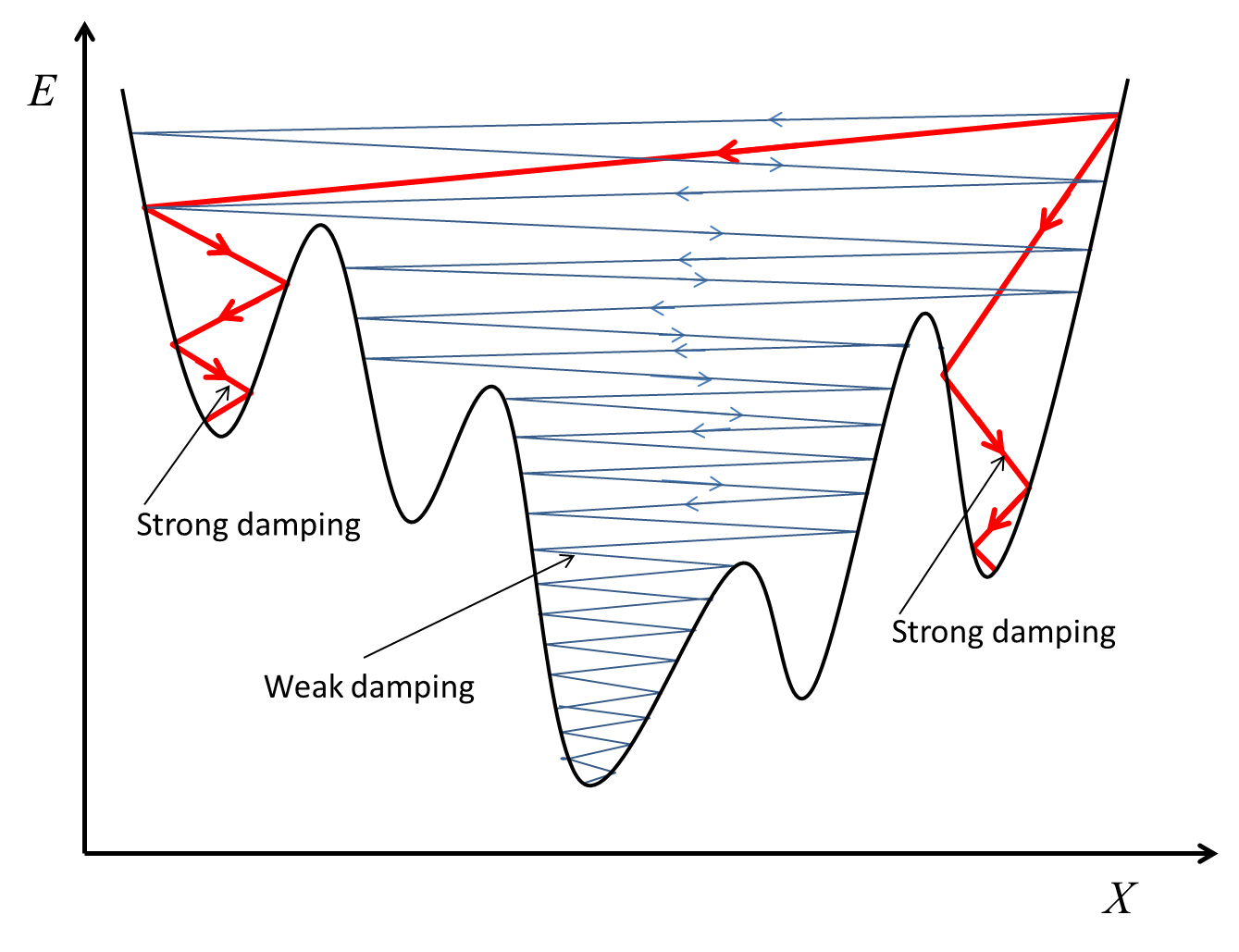}
\caption{Efficiency of the weak-damping method for glassy systems. }
\label{Fig:Energy_minimization}
\end{figure}

\begin{figure}
\begin{centering}
\includegraphics[width=7cm]{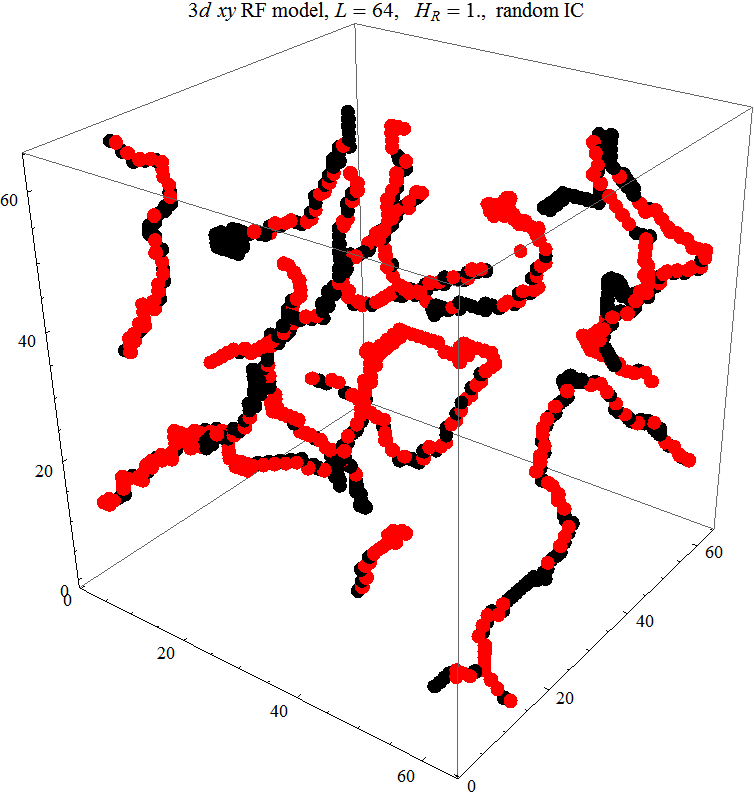}
\par\end{centering}
\caption{Pinned vortex loops obtained by relaxation from random
initial orientations of spins with $h\equiv H_R = 1$.\label{fig:vortex-loops}}
\end{figure}

In early 1980s, however, renormalization group results have
appeared, most noticeably by Cardy and Ostlund
\cite{Cardy-PRB1982} and by Villain and Fernandez
\cite{Villain-ZPB1984}, that questioned the validity of the LIM
theory at $R \geq R_f$. Application of scaling and replica-symmetry
breaking arguments by Nattermann \cite{Nattermann}, Korshunov
\cite{Korshunov-PRB1993} and by Giamarchi and Le Doussal
\cite{Giamarchi} yielded $\langle[\phi({\bf r}_1) - \phi({\bf
r}_2)]^2\rangle = A \ln|{\bf r}_1 - {\bf r}_2|$ at large distances, with $A$
depending on the dimensionality only.  This implies a universal power law decay of correlations,
$\sim 1/R$ according to Eq. (\ref{CF}), i.e., an ordering more robust against weak static randomness than
expected from the LIM theory. Such a {\em quasiordered} phase, presumed to be
vortex-free in spin systems and dislocation-free in Abrikosov
lattices, received the name of Bragg glass.

In parallel with analytical studies, the effect of static disorder
has been investigated by numerical methods. Early results on $1d$
\cite{DC-PRB1991} and $2d$ \cite{Dieny-PRB1990} systems with
quenched randomness have established strong non-equilibrium
effects, such as magnetic hysteresis and dependence on initial
conditions, as well as significant departure of the correlation
functions from the prediction of the LIM theory. Gingras and Huse
\cite{Gingras-Huse-PRB1996} attempted to test numerically the
existence of the vortex-free Bragg glass phase in $2d$ and
$3d$ random-field $xy$ spin model. While they found
some evidence of the expulsion of vortices below the critical
temperature, rapid freezing of spin dynamics prevented them from
making a definitive comparison with the Bragg glass theory. In a
follow-up on Ref. \cite{Gingras-Huse-PRB1996}, further
argument in favor of the Bragg glass phase was given by Fisher
\cite{Fisher-PRL1997} who analyzed energies of randomly pinned
dislocation loops. Defect-free models with relatively large random
field and random anisotropy have been studied numerically on small
lattices by Fisch \cite{Fisch}. At elevated temperatures the
numerical evidence of the power-law decay of correlations in a
$2d$ random-field $xy$ model has been recently obtained
by Perret et al. \cite{Perret-PRL2012}.

In spite of the large body of work, some fundamental questions
remain. Firstly, as in the situation addressed by the hairy ball
theorem \cite{HBT}, it is not obvious whether a vortex-free state
with zero magnetization due to the random
field can exist. Secondly, if the initial state contains vortices, it is not
clear whether they can escape pinning by disorder during the
relaxation process, so that the system evolves towards the Bragg
glass. Increased computational capabilities allow one to address
these questions, the second one being the main subject of this Letter.

Numerical method employed here combines finite rotation
(FR) and over-relaxation (OR) protocols. The FR update used in Ref. \cite{Dieny-PRB1990} rotates
each spin towards the direction of the local effective field,
${\bf H}_{i,{\rm eff}} = \sum_iJ_{ij}{\bf s}_j + {\bf h}_j + {\bf
H}$, while the OR update provides energy-conserving spin flips:
${\bf s}_i \rightarrow 2({\bf s}_i\cdot {\bf H}_{i,{\rm eff}}){\bf
H}_{i,{\rm eff}}/H_{i,{\rm eff}}^2 - {\bf s}_i$. Whereas the FR
method is searching for the energy minimum, the OR method is
searching for the entropy maximum. The FR and OR updates are
applied with the probabilities $\alpha$ and $1 - \alpha$
respectively, where $\alpha$ plays the role of a relaxation
constant. The fastest convergence was observed for $0.01\leq
\alpha \leq 0.1$ that physically corresponds to slow cooling.
Efficiency of the combined weak damping method for glassy systems is shown in
Fig. \ref{Fig:Energy_minimization}, assuming that deeper minima have broader basins of attraction.

Our Wolfram Mathematica program using
compilation and parallelization is comparable in speed with
programs written in Fortran and C and allows to relax systems of up to one
billion spins on our 96 GB RAM workstation. In numerics we use $J =a=s=1$
and $H_R$ instead of $h$.
All results presented below were obtained with periodic boundary conditions.

Random initial orientations of spins result in the evolution
towards the vortex-glass (VG) state. Vortices in a $3d$ lattice
form vortex loops shown in Fig. \ref{fig:vortex-loops}.
It is clear from the numerical work that the vortex-free Bragg
glass cannot be achieved because of the freezing of the system
into a VG state virtually at any temperature below the temperature
of local ferromagnetic ordering.

\begin{figure}
\begin{centering}
\includegraphics[width=8cm]{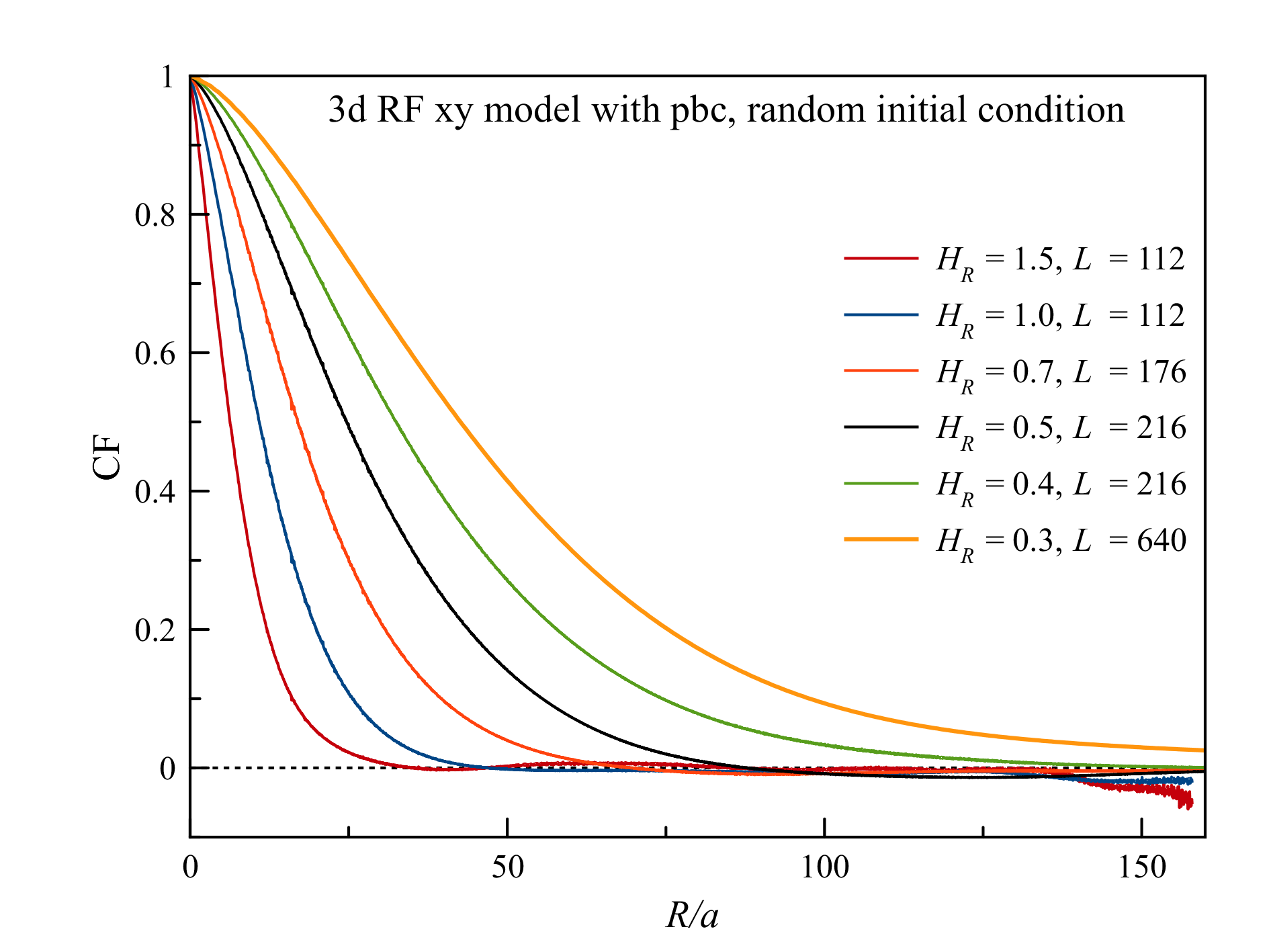}
\par\end{centering}
\caption{Spin-spin correlation function in the vortex-glass state
for different values of the random field $h\equiv H_R$. Smaller
$H_R$ require a larger system size $L$. \label{fig:CF}}
\end{figure}

\begin{figure}
\begin{centering}
\includegraphics[width=8cm]{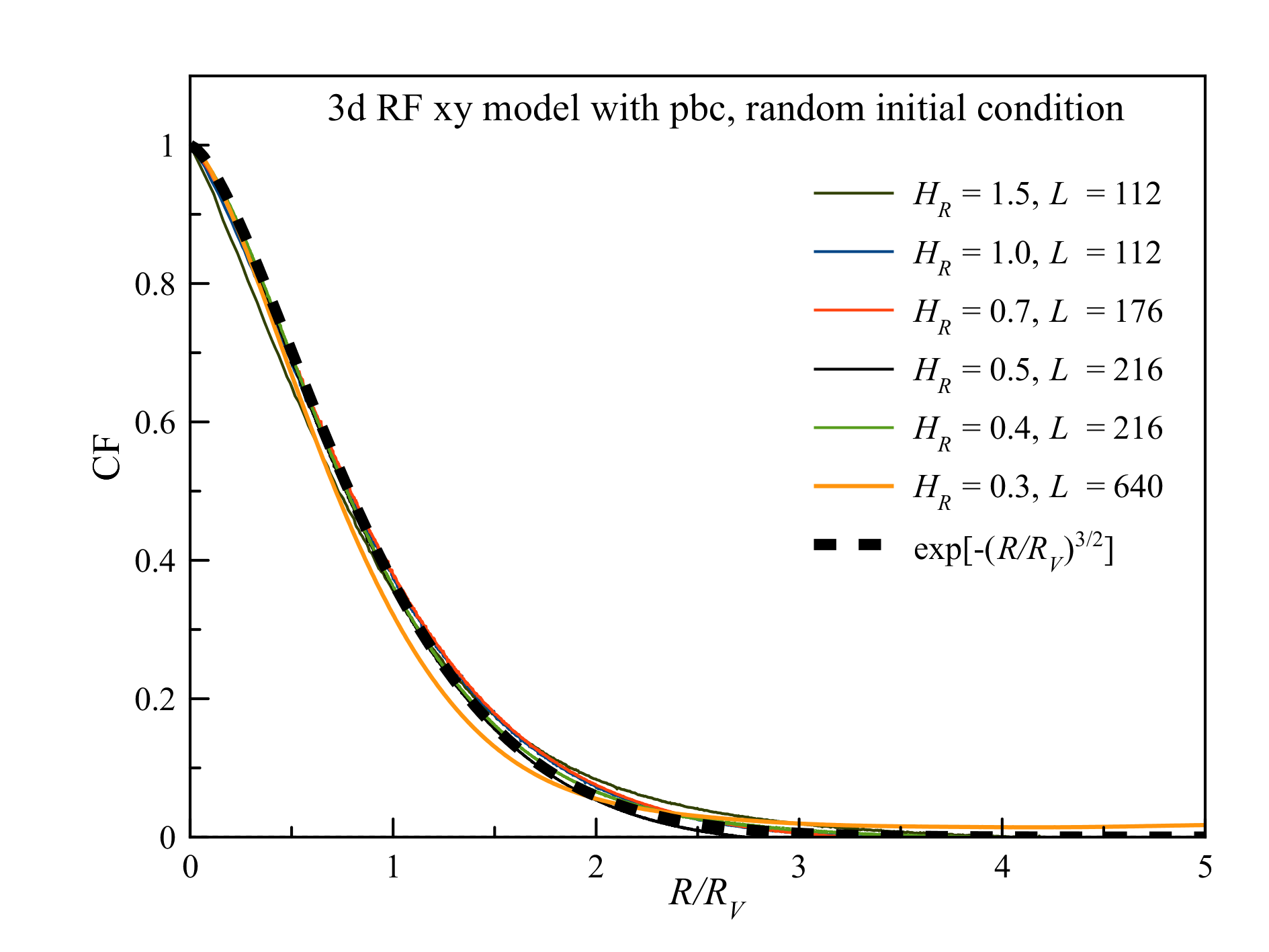}
\par\end{centering}
\caption{Spin-spin correlation function in the vortex-glass state,
scaled with the help of Eq.\ (\ref{Rv}). \label{fig:CF-scaled}}
\end{figure}

Spin correlation function (CF) in the VG state for different
amplitude of the random field, $h$, is shown in Fig. \ref{fig:CF}.
As Fig. \ref{fig:CF-scaled} demonstrates, the curves in Fig.
\ref{fig:CF} can be  scaled by
\begin{equation}\label{Rv}
\langle{\bf s}({\bf r}_1)\cdot{\bf s}({\bf r}_2)\rangle =
s^2e^{-(|{\bf r}_1 - {\bf r}_2|/R_v)^{3/2}}, \quad R_v \approx 14
\left(J/h\right)^{1.2}.
\end{equation}
In this stretched-exponential CF, the dependence of the correlation length $R_v$ on $h$ is much
weaker than that of $R_f$ provided by the LIM theory in
$3d$, Eq. (\ref{LIM-Rf}). Thus at small $h$ one has $R_v \ll R_f$. Qualitatively, this
result can be understood along the following lines. $R_v$ can be
roughly associated with the average distance between vortex lines.
The exchange energy per spin of the line scales as
$(Js^2/R_v^2)\ln R_v$. The vortex line is adjusting to the pinning
potential in every $xy$ plane independently. The pinning energy
per spin in each plane can be estimated along the lines of the LIM
argument in $2d$: $-hs\sqrt{R_v^2}/R_v^2 \sim -hs/R_v$.
Minimization of the sum of these two energies gives $R_v \sim
(Js/h)\ln R_v$, which may explain the power $1.2$ in Eq.\
(\ref{Rv}). It is closer to $R_f \sim (Js/h)$ in
$2d$ than to $R_f \sim (Js/h)^2$ in $3d$, because vortices are essentially two-dimensional objects.

\begin{figure}
\begin{centering}
\includegraphics[width=8cm]{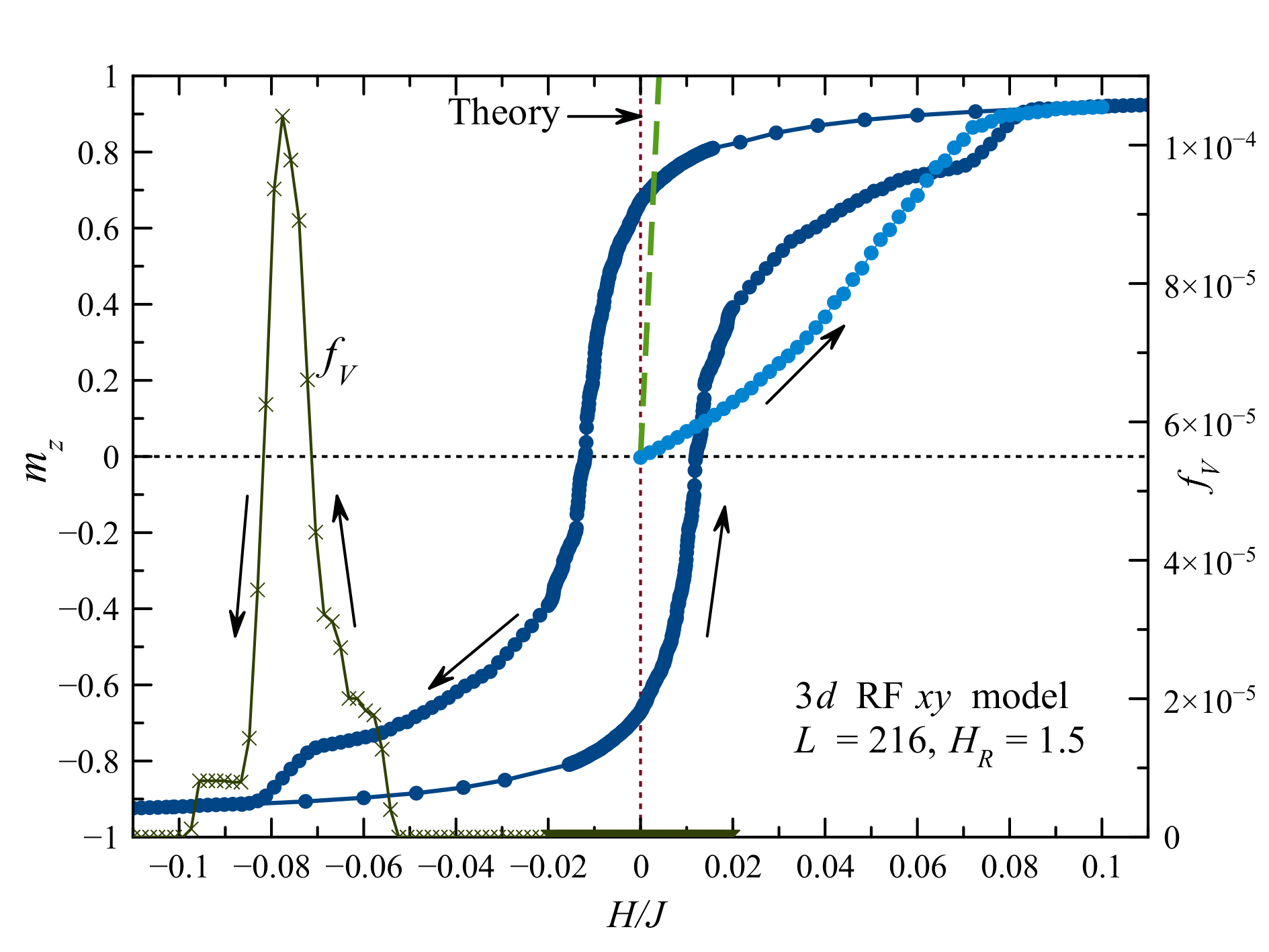}
\par\end{centering}
\vspace{-0.5cm}
\caption{Hysteresis curve for $3d$ random-field $xy$ model at $h =
1.5$. The initial magnetization curve of the VG state is shown by
the light blue curve. Green dashed line indicates the initial
slope in the LIM theory. Green solid line shows vorticity along
the hysteresis curve. \label{fig:hysteresis}}
\end{figure}

When placed in the external field, the random-field system exhibits
magnetic hysteresis. Magnetization curves at $h =
1.5$ are shown in Fig. \ref{fig:hysteresis}.

\begin{figure}
\begin{centering}
\includegraphics[width=7cm]{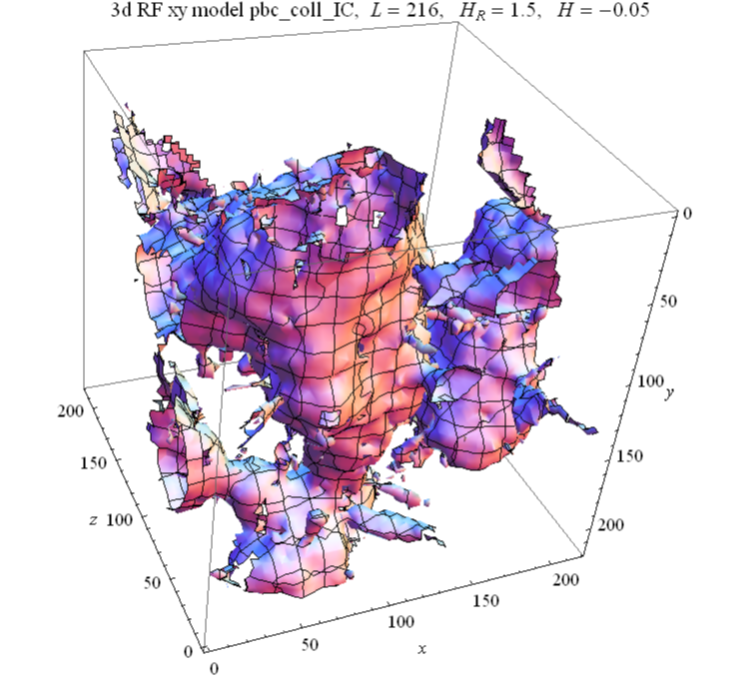}
\par\end{centering}
\vspace{-0.5cm}
\caption{Walls of spins opposite to the field emerge on decreasing
the magnetic field. \label{fig:walls}}
\end{figure}

The initial magnetization curve of the VG state, that is indicated
by the light blue line, corresponds to a rather hard magnet whose
properties are determined by strong pinning of the vortex lines in
the VG state. The initial susceptibility is much lower than
calculated within the LIM theory that disregards vortices, $\chi =
\frac{1}{2}(R_f/a)^2$. The latter correlates with the slope of the
hysteresis loop at $m_z =0$. On increasing the field vortices
are expelled from the sample at $H/J\approx 0.08$.

On decreasing the field, the lines do not re-enter the sample
until the field reaches a particular negative value. This can be seen from the green
vorticity curve in Fig. \ref{fig:hysteresis}. The latter has been
computed numerically by analyzing rotation of spin vectors along
any unitary square plaquette in the $xy$ planes \cite{Gingras-Huse-PRB1996}. Rotation of spins
by $\pm 2\pi$ indicates presence of a vortex or an antivortex.
Vorticity $f_V$ is defined as the fraction of plaquettes that
contain singularities of either sign.

\begin{figure}
\begin{centering}
\includegraphics[width=7cm]{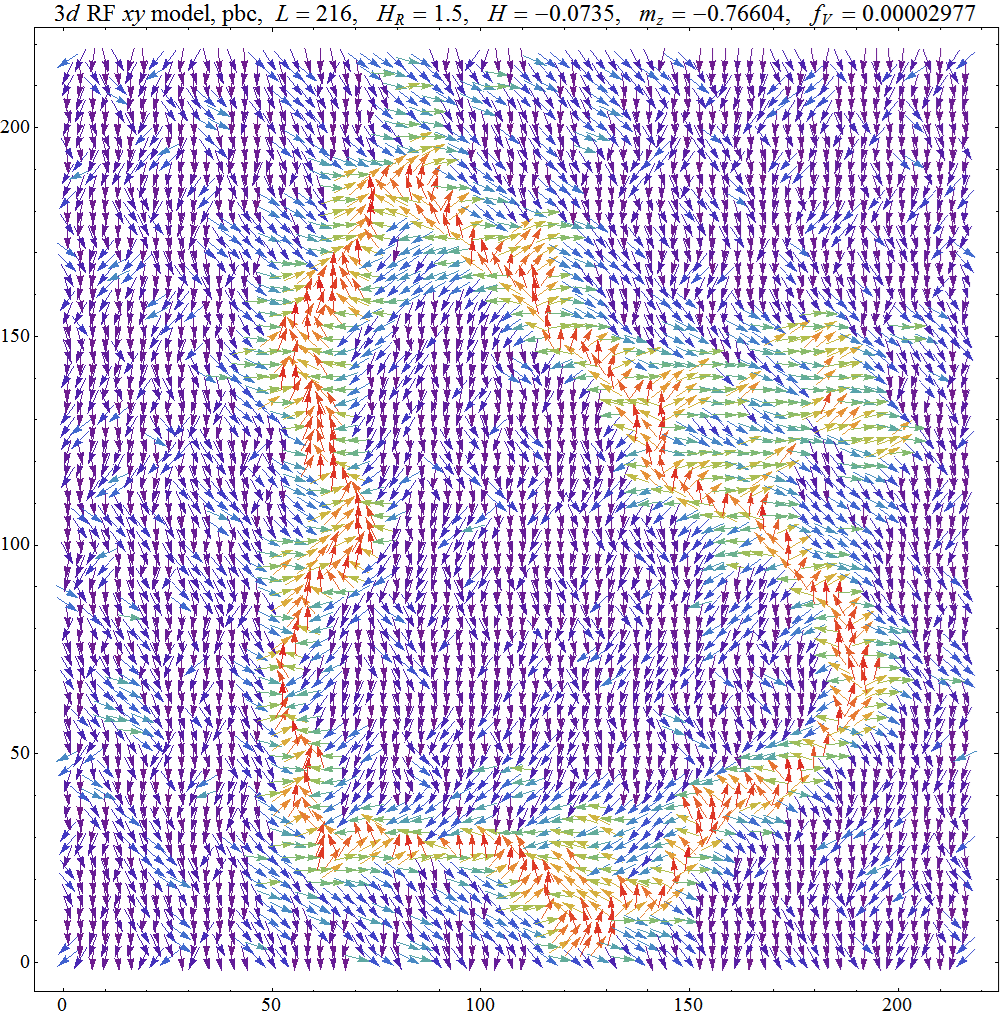}
\par\end{centering}
%
\caption{Cross-section of the sample showing walls of spins
(orange) opposite to the downward magnetic field.
\label{fig:SS-1}}
\end{figure}

Prior to vortices re-entering the sample, peculiar topological
objects appear in the system. These are walls of spins
opposite to the field, see Figs. \ref{fig:walls} and \ref{fig:SS-1}, that separate
regions where spins rotate in different directions towards the
direction of the field.

\begin{figure}
\begin{centering}
\includegraphics[width=7cm]{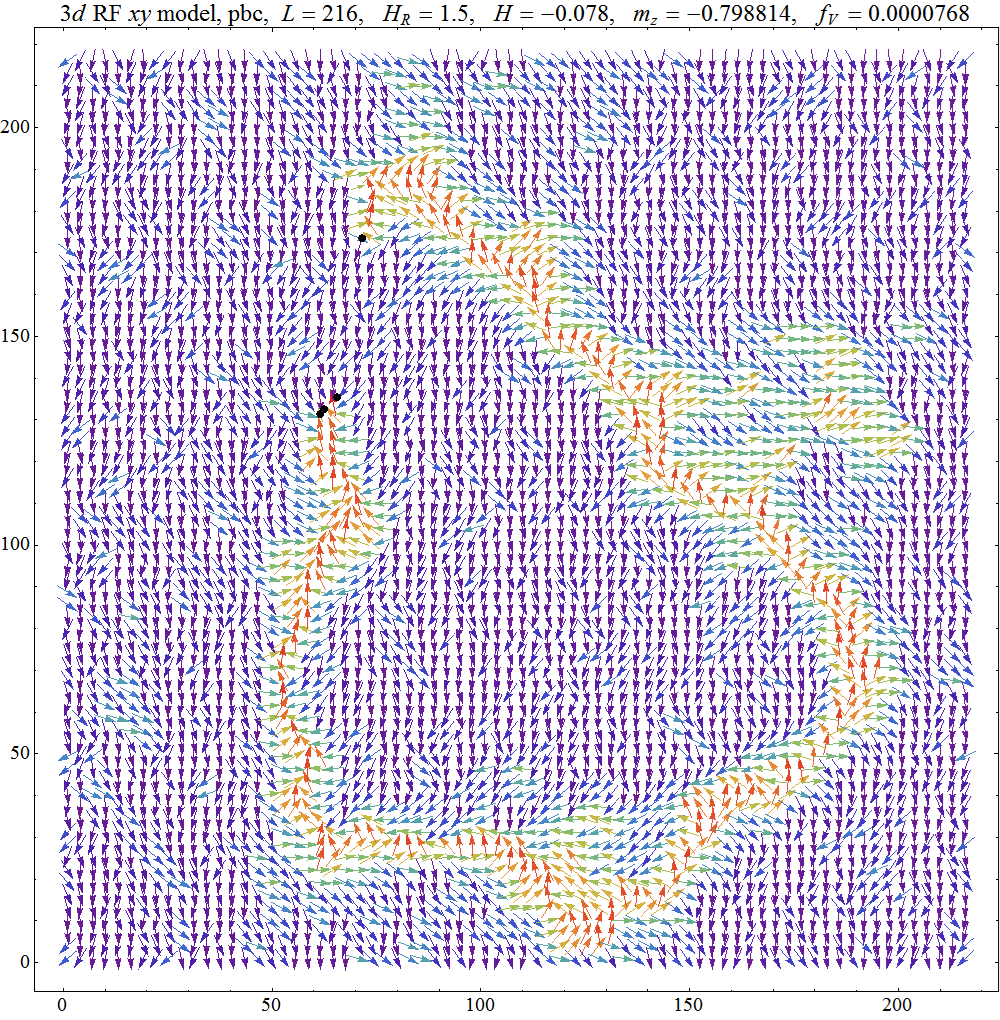}
\par\end{centering}
\caption{Cross-section of the wall after it develops a growing
crack bounded by the vortex loop (black points). \label{fig:SS-2}}
\end{figure}

Magnetization across the segment of the wall normal to, e.g., the
$x$ direction at $h = 0$ is given by the extremum of Eq.
(\ref{Ham-continuous}) in the class of functions $\phi = \phi(x)$:
$d^2\phi/dx^2 = (H/Js)\sin\phi$. Solution corresponds to the
rotation of the spin field by $2\pi$ as one moves across the wall,
$\phi = \pm 4 \arctan e^{ x/R_H}$, where $R_H = (Js/H)^{1/2}$.
Here $\pm$ correspond to the two possible directions of spin rotation in the wall.
 These walls are $3d$ counterparts of the
topological structures observed in a $2d$ random-anisotropy model in
Ref. \cite{Dieny-PRB1990}. In the absence of the random
field they would be driven out of the system or collapse as it reaches
equilibrium. Pinning of the walls by the random field makes them
stable. On increasing the
field strength, the rotation of the spin-field by $2\pi$ inside the walls
makes them increasingly weak regions against formation of vortex-antivortex
pairs at the atomic scale that changes the topology. As a result, the walls develop cracks bounded by vortex
loops. The loops then grow in size and eat the walls away. This
process is illustrated by snapshots of a cross-section of the wall  at two consecutive moments of time, Figs. \ref{fig:SS-1} and
\ref{fig:SS-2}.
Destruction of spin walls leads to the shoulder in the hysteresis curve seen at $H/J=-0.07$ in Fig. \ref{fig:hysteresis}.

In conclusion, our results show that studies of random field
models in the context of spin systems and flux lattices can hardly
ignore defects such as vortices and dislocations. This also
applies to Josephson junctions arrays with disorder \cite{SI}
where proximity to a bulk superconductor mimics the effect of the
external field. In our slow-cooling numerical experiment, the system with initially
random spin orientations freezes into a vortex glass state that exhibits
a faster decay of spin-spin correlations than predicted by the theory that does not take vortices into account.
Irreversible behavior
manifested by hysteresis, emerging wall-like topological structures,
and non-monotonic field dependence of vorticity is observed in the
external magnetic field.

This work has been supported by the Department of Energy through grant
No. DE-FG02-93ER45487.


\begin{thebibliography}{10}
\bibitem{Larkin-JETP1970} A. I. Larkin, Sov. Phys. JETP {\bf 31}, 784 (1970).

\bibitem{Imry-Ma-PRL1975} Y. Imry and S.-k. Ma, Phys. Rev. Lett. {\bf 35}, 1399 (1975).

\bibitem{RA} R. Pelcovits, E. Pytte, and J. Rudnick, Phys. Rev. Lett. {\bf 40}, 476 (1978);
A. Aharony and E. Pytte, Phys. Rev. Lett. {\bf 45}, 1583 (1980);
E. M. Chudnovsky, W. M. Saslow and R. A. Serota, Phys. Rev. B {\bf
33}, 251 (1986).

\bibitem{Aizenman-Wehr} M. Aizenman and J. Wehr, Phys. Rev. Lett. {\bf 62}, 2503 (1989).

\bibitem{Fisher-PRB1985} D. S. Fisher, Phys.Rev. B {\bf 31},7233
(1985).

\bibitem{Blatter-RMP1994} G. Blatter, M. V. Feigel'man, V. B. Geshkenbein,
A.I. Larkin, and V. M. Vinokur, Rev. Mod. Phys. {\bf 66}, 1125
(1994).

\bibitem{Cardy-PRB1982} J. L. Cardy and S. Ostlund, Phys. Rev. B
{\bf 25}, 6899 (1982).

\bibitem{Villain-ZPB1984} J. Villain and J. F. Fernandez, Z. Phys.
B - Condens. Matter {\bf 54}, 139 (1984).

\bibitem{Nattermann} T. Nattermann, Phys. Rev. Lett. {\bf 64},
2454 (1990); J. Kierfield, T. Nattermann, and T. Hwa, Phys. Rev. B
{\bf 55}, 626 (1997).

\bibitem{Korshunov-PRB1993} S. E. Korshunov, Phys. Rev. B {\bf
48}, 3969 (1993).

\bibitem{Giamarchi} T. Giamarchi and P. Le Doussal, Phys. Rev.
Lett. {\bf 72}, 1530 (1994); T. Giamarchi and P. Le Doussal, Phys.
Rev. B {\bf 52}, 1242 (1995); P. Le Doussal, Phys. Rev. Lett. {\bf
96}, 235702 (2006); P. Le Doussal amd K. J. Wiese, Phys. Rev.
Lett. {\bf 96}, 197202 (2006); A. A. Middleton, P. Le Doussal, and
K. J. Wiese, Phys. Rev. Lett. {\bf 98}, 155701 (2007).

\bibitem{DC-PRB1991} R. Dickman and E. M. Chudnovsky, Phys. Rev. B
{\bf 44}, 4397 (1991).

\bibitem{Dieny-PRB1990} B. Dieny and B. Barbara, Phys. Rev. B {\bf
41}, 11549 (1990).

\bibitem{Gingras-Huse-PRB1996} M. J. P. Gingras and D. A. Huse,
Phys. Rev. B {\bf 53}, 15193 (1996).

\bibitem{Fisher-PRL1997} D. Fisher, Phys. Rev. Lett. {\bf 78},
1964 (1997).

\bibitem{Fisch} R. Fisch, Phys. Rev. B {\bf 52}, 12512 (1995);
ibid {\bf 55}, 8211 (1997); ibid {\bf 57}, 269 (1998); ibid {\bf
62}, 361 (2000); ibid {\bf 76}, 214435 (2007); ibid {\bf 79},
214429 (2009).

\bibitem{Perret-PRL2012}
A. Perret, Z. Ristivojevic, P. Le Doussal, G. Schehr, and K. J.
Wiese, Phys. Rev. Lett. {\bf 109}, 157205 (2012).

\bibitem{HBT} See, e.g., J. W. Milnor, {\it Topology From the
Differentiable Viewpoint} (The University Press of Virginia,
Charlotesville, 1965).

\bibitem{SI}
E. M. Chudnovsky, Phys. Rev. Lett. {\bf 103}, 137001 (2009).

\end{thebibliography}
\end{document}